\documentclass[11pt,a4paper]{article}
\usepackage{amssymb}
\usepackage{graphicx}
\usepackage{layout}
\usepackage{amssymb}
\begin{document}

\author{Guido Rizzi$^{\S}$  and Matteo Luca Ruggiero$^{\S,\ast}$
\\ \\
\small
%%EndAName
$^\S$ Dipartimento di Fisica, Politecnico di Torino,\\
\small Corso Duca degli Abruzzi 24, 10129 Torino, Italy \\
\small $^\ast$ INFN, Via P. Giuria 1, 10125 Torino, Italy\\ \\
 E-mail rizzi@polito.it, ruggierom@polito.it}
\title{Space geometry of rotating platforms: an operational approach}
\maketitle

\normalsize
\begin{abstract}
We study the space geometry of a rotating disk  both from a
theoretical and operational approach; in particular we give a
precise definition of the space of the disk, which is not clearly
defined in the literature. To this end we define an extended
3-space, which we call \textit{relative space}: it is recognized
as the only space having an actual physical meaning from an
operational point of view, and it is identified as the 'physical
space of the rotating platform'. Then, the geometry of the space
of the disk turns out to be non Euclidean, according to the early
Einstein's intuition; in particular the Born metric is recovered,
in a clear and self consistent context. Furthermore, the
relativistic kinematics reveals to be self consistent, and able to
solve the Ehrenfest's paradox  without any need of dynamical
considerations or \textit{ad hoc} assumptions.
\end{abstract}


\section{Introduction}\label{sec:intro}

It is a common belief that the conceptual foundations and the
experimental predictions of the special theory of relativity (SRT)
are accepted by everyone in a widespread agreement about the
theory. Actually, even in recent times, after one century of
relativity, some problems are still under discussion and the
debate about them is somewhat lively. One important topic is the
rotation of the reference frame which, contrary to the
translation, has an absolute character, and can be locally
measured by the Foucault's pendulum or by the Sagnac experiment.
This peculiarity of rotation, inherited by newtonian physics, is
difficult to understand in a relativistic context. Since the very
beginning of last century, those authors who were contrary to SRT
had found, in the relativistic approach to rotation, important
arguments against the self-consistency of the theory. In fact, in
1909 an internal contradiction in the SRT, applied to the case of
a rotating disk, was pointed out by Ehrenfest \cite{ehrenfest};
few years later, in 1913, a contradiction of SRT with experimental
data was pointed out by Sagnac \cite {sagnac}.

Since then, both the so-called 'Ehrenfest's paradox' and the
theoretical interpretation of the Sagnac effect had become topical
arguments of a discussion on the foundations of the SRT, which is
not closed yet, as the increasing number of recent contributions
confirms.

This paper deals with the 'Ehrenfest's paradox', which arises when
a circular platform, initially at rest, is set into rotation
around its symmetry axis. When the acceleration period finishes,
the disk keeps rotating with constant angular velocity $\omega $.
According to SRT, the rim of the circular platform does undergo
the standard Lorentz contraction, with respect to the inertial
observer at rest in the center of the disk, while the radius does
not. Hence the ratio of the circumference to the radius, as
measured by the inertial observer, should be less than $2\pi $,
thus violating the Euclidean geometry of the inertial frame.

After one hundred years of relativity, the problem of the rotating
disk does not seem to be solved yet; or more explicitly, there is
not a common agreement on its solution. This is confirmed by the
list of the authors who faced and tried to solve the Ehrenfest's
paradox, which we mention in the brief historical excursus of next
section.\\

Some of the given solutions will be discussed in details in the
following sections, in comparison to our approach, by which the
paradox will be solved on the bases of purely kinematical
arguments. We believe that lots of misunderstandings arose because
of the lack of proper definitions of some crucial concepts in the
theoretical apparatus needed to solve the paradox. In particular,
our approach will be based on (i) a precise definition of the
concept of ''space of the disk'', (ii) a precise choice of the
''standard rods'' used by the observers on the platform (which
ultimately agrees with the choice usually done in relativity). Of
course, the geometry of the space of the disk rests on the
assumptions (i), (ii). We shall see that the space of the disk
turns out to be non-Euclidean\footnote{The non null curvature of
the space of the disk has nothing to do with the spacetime
curvature which is and remains null (in each chart used to compute
it, and in each frame to which the chart is adapted) if
gravitation is not taken into account.}, and its metric coincides
with the one which is found in classic textbooks of relativity,
\textit{in spite of a shift of the context}. In fact, we shall
give a new definition of the ''space of the disk'', compatible
with an operational procedure of space and time measurements
performed by the observer on the platform. The projection
technique introduced by Cattaneo will be adopted to approach the
problem; this allows a straightforward description of the spatial
geometry of any reference frame. For a rotating reference frame
our results turn out to be, unexpectedly, the same as those known
in literature \textit{because of the underlying symmetries}. In
the general case (in particular non axis- symmetric and
non-stationary) no symmetries are present, and an approach similar
to the one proposed in this paper is mandatory to study the
spatial geometry of these frames.

\section{A little history of the Ehrenfest's Paradox}
\label{ssec:history}

\indent  \indent \setcounter{subsection}{1} \textbf{\thesubsection
\ Dynamical approaches to the paradox.}\label{ssec:dynapp}
 Soon after the publication of Ehrenfest's paper (1909),
Planck\cite{planck} pointed out that the paradox should be
investigated as an elasticity problem, questioning the possibility
of the application of Lorentz contraction to a body which is
brought from rest to uniform motion. Starting from similar ideas,
Lorentz\cite{lorentz} and Eddington\cite{eddington}, in the 20's,
deduced that both the radius and the circumference contract in the
same way; thus the paradox disappeared. However a more detailed
analysis of the problem was given, by means of a dynamical
treatment, by Clark\cite{clark} in 1947, who pointed out that
Lorentz and Eddington had overlooked that the velocity of
propagation of a dilation cannot exceed the speed of light. Taking
this into account, he showed that for material in which waves
travel with the fundamental velocity, no alteration of the radius
of the disk is
present\footnote{%
The same result was obtained 5 years before by Berenda
\cite{berenda}, using a purely kinematical approach.}. More
recently, in the 60's, Cavalleri\cite {cavalleri} tried to give a
dynamical solution of the Ehrenfest's paradox, claiming that it
''cannot be solved from a purely kinematical point of view''. We
believe that this statement is too radical, and we agree with
Phipps\cite{phipps1}, who observed that ''the fact that dynamics
can exist, without the foundation of logically consistent
kinematics, is absurd'', and concluded that ''kinematics is
foundational (logically preconditional) to
physics''.\footnote{%
Other dynamical aspects of the relativistic description are treated by Brotas%
\cite{brotas} and McCrea\cite{mccrea}. }\\

  \indent \setcounter{subsection}{2} \textbf{\thesubsection \
Kinematical approaches to the paradox.}\label{ssec:kinapp} By
means of purely kinematical considerations, Einstein
\cite{einstein} tried to avoid the paradox suggesting that
rotation must distort the Euclidean geometry of the platform, so
that the geometry of the inertial frame remains Euclidean.
Einstein was interested in the problem of the rotating disk just
as an euristic tool in order to investigate the possibility that
the geometry of the Minkowskian spacetime could be distorted by a
gravitational field. In the words of Stachel \cite{stachel},
Einstein's approach to the problem of the spatial geometry of the
rotating disk ''showed that a gravitational field (here equivalent
to the centrifugal field) causes non Euclidean arrangement of
measuring rods, and thus compelled a generalization of Euclidean
space''. So  Einstein's treatment of the rotating disk seems to
provide a ''missing link'' in the chain of reasoning from the SRT
towards a relativistic theory of gravitation (GRT), based on the
idea that the geometry of the familiar Minkowskian spacetime of
the SRT must be deformed
by a gravitational field.\footnote{%
Let us comment that this argument, though important from a
historical and euristic point of view, is questionable, since it
relies upon the application of the principle of equivalence, whose
validity is strictly local, to the global space of the disk.}.

An early alternative attempt to solve the paradox was made in 1910
by Stead-Donaldson\cite{stedon}, and later taken up by Ives\cite{ives}, Eagle%
\cite{eagle}, Galli\cite{galli}. This attempt consists in
supposing that the surface of the disk bends because of rotation,
assuming the shape of a paraboloid of revolution. This hypotesis
explains in a very simple way the deformation of the Euclidean
geometry of the disk: however, it neglects the kinematical
constraints imposed on the motion of the disk (i.e. rigid
2-dimensional motion), and moreover introduces further
difficulties. In particular, a non symmetric deformation with
respect to the plane of the non-rotating disk should determine a
screw sense in space, thus violating spatial parity \textit{in a
purely kinematical context}.

In order to avoid speeds of the points of the disk that would
exceed the velocity of light at great distance from the axis of
rotation (as observed in the inertial frame), some authors, like
Hill\cite{hill} in 1946, suggested that the speed-distance law
should be non linear. As a consequence, Hill claimed that the
problem of the space geometry of the rotating disk is a
''ill-defined question'', since it strictly depends on the way the
disk is set into rotation. However this hypothesis was ruled out
 by Rosen\cite{rosen} in 1947.

The first important contribution to the study of the kinematical
aspects of the problem was given in 1942 by Berenda\cite{berenda},
with particular attention to the meaning of the spatial geometry
of the disk. Similar
considerations can be found in Rosen\cite{rosen}, Arzeli\`{e}s\cite{arzelies}%
, Landau-Lifshits\cite{landau}, M\o ller\cite{moller}, Gr\o
n\cite{gron1}, \cite{gron2}, \cite{gron3}. Also
Weber\cite{weber1}, Dieks\cite{dieks}, Anandan\cite{anandan},
Rizzi-Tartaglia\cite{rizzi1}, Bergia-Guidone \cite{bergia}, were
interested in the kinematics of the rotating disk, focusing on the
relativistic interpretation of the Sagnac effect.

As mentioned in the Introduction, the Sagnac effect was
interpreted as a disproval of SRT, both during the early years of
SRT by Sagnac himself (1913), and more recently,
in the 90's,  by Selleri\cite{selleri1},\cite{selleri2}, Croca and Selleri%
\cite{croca}, Goy and Selleri\cite{goy}, Vigier\cite{vigier},
Anastasovski \textit{et al.}\cite{anas}. However this claim is
uncorrect, as showed in the papers quoted
(\cite{weber1}-\cite{bergia}) and more recently (2001) by
Rodrigues and Sharif\cite{rodrigues}: \textit{the Sagnac effect
can be explained in the framework of the SRT as a pure
desynchronization effect}. Let us briefly recall the most
important experimental clue of this claim. As a matter of fact,
any pair of luminal or material objects (electromagnetic or matter
waves), travelling in opposite directions along the rim, with the
same relative velocity, take different times for the total round
trip; but the time difference \textit{between any pair of
travelling objects} is always the same, and exactly coincides with
the synchronization gap predicted by the SRT for
non-time-orthogonal physical frames.

Some authors tried to solve the Ehrenfest's paradox proposing
theories which are alternative to SRT, removing the conceptual
bases of the paradox itself. In particular, Klauber in 1998
\cite{klauber},\cite{klauber2} proposed a ''New Theory of Rotating
Frames'' in which the Lorentz contraction takes place only in the
case of translation, but not in the case of rotation.

More recently, Tartaglia\cite{tartaglia} agreed with the
assumption ''no Lorentz contraction for rotating frames''; however
he stressed that this assumption is compatible with the SRT,
solving the paradox in a strictly relativistic context and from a
pure kinematical point of view. His claim depends on the
assumption that, in the case of rotations, the relativity of
simultaneity is not essential for the space measurement processes
of the observer at rest on the platform. In his words: ''the
Lorentz contraction is a manifestation of the relativity of
simultaneity in different inertial
frames: \textit{when no synchronization is needed, no contraction appears}%
''. As a consequence, he concluded that the length of an
infinitesimal standard rod on the platform cannot be altered by
rotation, and the space of the disk must be flat.

Similar conclusions were reached in the 70's by
Gr\"{u}nbaum-Janis\cite {grjan} and Strauss\cite{strauss}, on the
ground of different assumptions. The former claimed that the
length of the periphery, as measured in the rotating frame, is
equal to its rest length; the latter asserted that both the
measuring rods along the circumference and the distances they are
supposed to measure (i.e. the length of the circumference) are
Lorentz-contracted with respect to the inertial frame: the two
effects cancel each other so that the space of the disk turns out
to be Euclidean. A first reply to these authors, which is
compatible with our conclusions, was given by Gr\o
n\cite{gron1},\cite{gron2} in the late 70's. He showed that the
construction of an acceleration program, which keeps the rest
length of every element of the disk periphery constant, during and
after the period of angular acceleration, is a kinematical
impossibility in the SRT context. As far as we know, he has been
the first author who stressed the different behavior of standard
rods and elements of the circumference: the proper length of the
former is not affected by accelerations; the length of the latter
increases during the acceleration period, and then remain
increased when acceleration finishes.\newline

These are just the most outstanding attempts of solving, or only
facing, the Ehrenfest's paradox, and they prove how pervading the
problem is throughout the decades.

\section{The Ehrenfest's paradox and its solutions}\label{sec:para}

The formulation of the paradox in the words of Ehrenfest is the following:%
\newline

"...Let R be the radius appearing to the stationary [i.e.
inertial] observer, during its motion [R' is the radius observed
in the rest frame of the disk]. Then R  must satisfy two
conditions that are contradictory to each other:\newline
\newline
\indent (a) \ \ The circumference of the cylinder must show a
contraction relative to the rest state, $2 \pi R < 2 \pi
R^{\prime}$, since each element
of the circumference moves in its own direction with instantaneous speed $%
\omega R$. \newline
\newline
\indent (b) If one consider an element of a radius, its
instantaneous velocity is perpendicular to its length; thus, an
element of the radius
cannot show a contraction with respect to the rest state. Therefore $%
R=R^{\prime}$...".\newline
\newline
Then a contradiction arises, which is the core of the
paradox.\newline Ehrenfest pointed out the apparent inconsistence
of the kinematics of bodies which are rigid, according to the
definition of rigidity given by Born (see below).\newline Despite
the great number of attempts, the underlying strategies of
solutions are basically these:

\begin{itemize}
\item[(s1)]  lengths do not contract in the case of the rotating disk $%
\Longrightarrow 2\pi R^{\prime }=L^{\prime }=L=2\pi R$;\\
\textit{(f.i. Klauber, Gr\"{u}nbaum-Janis, Tartaglia)}
\item[(s2)]  both the radius and the circumference contract, so that their
ratio remains $2\pi $ $\Longrightarrow \frac{L^{\prime }}{R^{\prime }}=2\pi =%
\frac{L}{R}$; \\\textit{(f.i. Lorentz, Eddington)}

\item[(s3)] rods along the rim do contract, while the
circumference does not; neither the rods along the radius nor the
radius itself do contract. As a consequence the space of the disk
is not Euclidean: $\Longrightarrow 2\pi R=L<L^{\prime }=\gamma
2\pi R^{\prime }$, where $\gamma $ is the Lorentz factor;\\
\textit{(f.i. Einstein, Berenda, Rosen, M\o ller, Landau-Lifshits,
Arzeli\`{e}s, Dieks, Weber, Gr\o n)}

\item[(s4)] both the rods along the rim and the circumference
contract; neither the rods along the radius nor the radius itself
do contract; as a consequence the surface of the disk bends,
because of rotation\\ \textit{(f.i. Stead-Donaldson, Ives, Eagle,
Galli )}.
\end{itemize}

In this paper we shall introduce the concept of "relative space"
that is the mathematical model defining the physical space of the
disk, which in the literature  is not clearly defined. This fact
explains, in part, the differences in the results found by
different authors. The mathematical details of next section are
needed to give a proper definition of this fundamental concept.

\section{Cattaneo's projection technique}\label{sec:cattaneo}

For our purposes, it is important to define correctly the
properties of the physical frames with respect to which we
describe the measurement processes. Although we remain in a
special relativistic context, we shall adopt the most general
description, which takes into account non-inertial frames (f.i.
rotating frames) in SRT, and arbitrary frames in GRT.

\indent \setcounter{subsection}{1} \textbf{\thesubsection \
}\label{ssec:c1} The physical spacetime is a (pseudo)riemannian manifold $%
\mathcal{M}^{4}$, that is a pair $\left(
\mathcal{M},\mathbf{g}\right)$, where $\mathcal{M}$ is a connected 4-dimensional Haussdorf manifold and $%
\mathbf{g}$ is the metric tensor\footnote{%
The riemannian structure implies that $\mathcal{M}$ is endowed
with an affine connection compatible with the metric, i.e. the
standard Levi-Civita connection.}. Let the signature of the
manifold be $(1,-1,-1,-1)$. Suitably differentiability condition,
on both $\mathcal{M}$ and $\mathbf{g}$, are assumed.

\indent \setcounter{subsection}{2} \textbf{\thesubsection \
}\label{ssec:c2}A physical reference frame  is a time-like
congruence
 $\Gamma $: the set of the world lines of the test-particles
constituting the ''reference fluid''\footnote{The concept of
'congruence' refers to a set of word lines filling the manifold,
or some part of it, smoothly, continuously and without
intersecting.\\ The concept of 'reference fluid' is an obvious
generalization of the 'reference solid' which can be used in flat
spacetime, when the test particles constitute a global inertial
frame. In this case, their relative distance remains constant and
they evolve as a rigid frame.
\par
However: (i) in GRT  test particles can be subject to a
gravitational field (curvature of spacetime); (ii) in SRT test
particles can be subject to an acceleration field. In both cases,
global inertiality is lost and tidal effects arise, causing a
variation of the distance between them. So we must speak of
"reference fluid", dropping the compelling request of classical
rigidity.}. The congruence $\Gamma $ is identified by the field of
unit vectors tangent to its world lines. Briefly speaking, the
congruence is the (history of the) physical frame or the reference
fluid (they are synonymous).\newline

\indent \setcounter{subsection}{3} \textbf{\thesubsection \
}\label{ssec:c3}Let $\{x^{\mu }\}=(x^{{0}},x^{1},x^{2},x^{3})$ be
a system of coordinates in the neighborhood of a point $p\in
\mathcal{M}$; these coordinates are said to be \textit{admissible}
(with respect to the
congruence $\Gamma $) when\footnote{%
Greek indices run from 0 to 3, Latin indices run from 1 to 3.}
\begin{equation}
g_{{00}}>0\ \ \ \ g_{ij }dx^{i }dx^{j }<0 \label{eq:admiss}
\end{equation}
Thus the coordinates $x^{{0}}=var$ can be seen as describing the
world lines of the $\infty ^{3}$ particles of the reference
fluid.\\

\indent \setcounter{subsection}{4} \textbf{\thesubsection \
}\label{ssec:c4}When a reference frame has been chosen, together
with a set of admissible coordinates, the most general coordinates
transformation which does not change the physical frame, i.e. the
congruence $\Gamma $, has the form
\cite{moller},\cite{gron1},\cite{cattaneo}:
\begin{equation}
\left\{
\begin{array}{c}
x^{\prime }{}^{{0}}=x^{\prime }{}^{{0}}(x^{{0}},x^{1},x^{2},x^{3}) \\
x^{\prime }{}^{i}=x^{\prime }{}^{i}(x^{1},x^{2},x^{3})
\end{array}
\right.  \label{eq:gauge_trans}
\end{equation}
with the additional condition $\partial x^{\prime 0}/\partial
x^{{0}}>0$, which  ensures that the  change of time
parameterization does not change the arrow of time. The
coordinates transformation (\ref
{eq:gauge_trans}) is said to be \textit{internal to the physical frame} $%
\Gamma $, or more simply  \textit{internal gauge
transformation}.\\

\indent \setcounter{subsection}{5} \textbf{\thesubsection \
}\label{ssec:c5}An ``observable'' physical quantity is in general
frame-dependent, but its physical meaning requires that it cannot
depend on the particular parameterization of the physical frame:
in brief it cannot be gauge-dependent. Then a problem arises. In
the mathematical model of GRT, physical quantities are expressed
by absolute entities\footnote{ `Absolute' means `independent of
any reference frame'.}, such as world tensors, and physical laws,
according to the covariance principle, are just relations among
these entities. So, given a reference frame, how do we relate
these absolute quantities to the relative, i.e.
reference-dependent, ones? And how do we relate world equations to
reference-dependent ones? In other words: how do we relate, by a
suitable 3+1 splitting, the mathematical model of spacetime to the
observable quantities which are relative to a reference frame?\\

\indent \setcounter{subsection}{6} \textbf{\thesubsection \
}\label{ssec:c6} \newcounter{CiSei}\setcounter{CiSei}{6} In order
to do that, we shall use the \textit{projection technique}
developed by Cattaneo  \cite{cattaneo} \cite{catt1}, \cite{catt2}, \cite{catt3}, \cite{catt4}%
. Let \mbox{\boldmath $\gamma$}$(x)$ be the field of unit vectors
tangent to the world lines of the congruence $\Gamma $. Given a
time-like congruence $\Gamma $ it is always possible to choose a
system of admissible
coordinate so that the lines $x^{0}=var$ coincide with the lines of $\Gamma $%
; in this case, such coordinates are said to be \textit{`adapted
to the physical frame'} defined by the congruence $\Gamma $.

Being $g_{\mu \nu }\gamma ^{\mu }\gamma ^{\nu }=1$, we get
\begin{equation}
\left\{
\begin{array}{c}
\gamma ^{{0}}=\frac{1}{\sqrt{g_{{00}}}} \\
\gamma ^{i}=0
\end{array}
\right. \;\;\;\;\;\;\;\;\;\;\;\;\left\{
\begin{array}{c}
\gamma _{0}=\sqrt{g_{{00}}} \\
\gamma _{i}=g_{i{0}}\gamma ^{{0}}
\end{array}
\right.  \label{eq:gammas}
\end{equation}

In each point $p\in \mathcal{M}$, the tangent space $T_{p}$ can be
split into the direct sum of two subspaces: $\Theta _{p}$, spanned
by $\gamma ^{\alpha }$, which we shall call \textit{local time
direction} of the given frame, and $\Sigma _{p}$, the
3-dimensional subspace which is supplementary (orthogonal) with
respect to $\Theta_{p}$; $\Sigma _{p}$ is called \textit{local
space platform} of the given frame. So the tangent space can be
written as the direct sum
\begin{equation}
T_{p}=\Theta _{p}\oplus \Sigma _{p}  \label{eq:tangsum}
\end{equation}

A vector $\mathbf{v}\in T_{p}$ can be projected onto $\Theta _{p}$ and $%
\Sigma _{p}$ using the \textit{time projector} $\gamma _{\mu }\gamma _{\nu }$
and the\textit{\ space projector }$\gamma _{\mu \nu }\doteq g_{\mu \nu
}-\gamma _{\mu }\gamma _{\nu }$:
\begin{equation}
\left\{
\begin{array}{rclll}
\bar{v}_{\mu } & = & P_{\Theta }\left( \,\mathbf{v}\right) &
\doteq &
\gamma _{\mu }\gamma _{\nu }v^{\nu } \\
\widetilde{v}_{\mu } & = & P_{\Sigma }\left( \,\mathbf{v}\right) &
\doteq & v_{\mu }-v^{\nu }\gamma _{\nu }\gamma _{\mu }=\left(
g_{\mu \nu }-\gamma _{\mu }\gamma _{\nu }\right) v^{\nu }=\gamma
_{\mu \nu }v^{\nu }
\end{array}
\right.  \label{eq:DefSplitComp}
\end{equation}

\textbf{Notation} The superscripts $^{-},^{\sim }$ denote
respectively a \textit{time vector} and a \textit{space vector},
or more generally, a \textit{time tensor }and\textit{\ a space
tensor} (see below).\\

Equation (\ref{eq:DefSplitComp}) defines the \textit{natural
splitting} of a vector. The tensors $\gamma _{\mu }\gamma _{\nu }$
and $\gamma _{\mu \nu }$ are called \textit{time metric tensor}
and \textit{space metric tensor}, respectively. In particular, for
each vector $\mathbf{v}$ it is possible to define a `time norm'
$\left\| \,\mathbf{v\,}\right\| _{\Theta }$ and a `space norm'
$\left\| \,\mathbf{v\,}\right\| _{\Sigma }$ as follows:
\begin{equation}
\left\| \,\mathbf{v}\,\right\| _{\Theta } \doteq \bar{v}_{\rho
}\bar{v}^{\rho }=\gamma _{\rho }\gamma _{\nu }v^{\nu }\left(
\gamma ^{\rho }\gamma _{\mu }v^{\mu }\right) =\gamma _{\mu }\gamma
_{\nu }v^{\mu }v^{\nu }=(\gamma _{\mu }v^{\mu })^{2} \geq 0
\label{eq:normtemp}
\end{equation}
\begin{equation}
\left\| \,\mathbf{v}\,\right\| _{\Sigma } \doteq \widetilde{v}_{\nu }\widetilde{v}%
^{\nu }= \gamma _{\mu \nu }v^{\mu }(v^{\nu }-\gamma _{\eta }\gamma
^{\nu }v^{\eta })=\gamma _{\mu \nu }v^{\mu }v^{\nu } \leq 0
\label{eq:normspaz}
\end{equation}

For a tensor field $T\in T_{p}$, every index can be projected onto $\Theta
_{p}$ and $\Sigma _{p}$ by means of the projectors defined before:
\begin{equation}
P_{\Theta }(T_{...\mu ...}) \doteq \gamma _{\mu }\gamma _{\nu
}T^{...\nu ...}\ \ \ P_{\Sigma }(T_{...\mu ...}) \doteq \gamma
_{\mu \nu }T^{...\nu ...}  \label{eq:protn}
\end{equation}

A tensor field of order two can be split in the sum of four tensor
\begin{equation}
\begin{array}{rclcrcl}
P_{\Sigma \Sigma }\,\left( \,{t}_{\mu \nu }\,\right) & \doteq &
\gamma _{\mu \rho
}\gamma _{\nu \eta }t^{\rho \eta } & \;\quad & P_{\Sigma \Theta }\,\left( \,{%
t}_{\mu \nu }\,\right) & \doteq & \gamma _{\mu \rho }\gamma _{\nu
}\gamma _{\eta
}t^{\rho \eta } \\
P_{\Theta \Sigma }\,\left( \,{t}_{\mu \nu }\,\right) & \doteq &
\gamma _{\mu }\gamma _{\rho }\gamma _{\nu \eta }t^{\rho \eta } &
\quad & P_{\Theta \Theta }\,\left( \,{t}_{\mu \nu }\,\right) &
\doteq & \gamma _{\mu }\gamma _{\nu }\gamma _{\rho }\gamma _{\eta
}t^{\rho \eta }
\end{array}
\label{eq:ProiettoriR2}
\end{equation}
belonging to four orthogonal subspaces
\begin{equation}
T_{p}\otimes T_{p}=(\Sigma _{p}\otimes \Sigma _{p}) \oplus (\Sigma
_{p}\otimes \Theta _{p}) \oplus (\Theta _{p}\otimes \Sigma _{p})
\oplus (\Theta _{p}\otimes \Theta _{p} )  \label{eq:Sottospazi}
\end{equation}

In particular, every tensor belonging entirely to $\Sigma _{p}\otimes \Sigma
_{p}$ is called a \textit{space tensor} and every tensor belonging to $%
\Theta _{p}\otimes \Theta _{p}$ is called a \textit{time tensor}.
Of course, these entities have a tensorial behavior only with
respect to the group of the coordinates transformation
(\ref{eq:gauge_trans}). It is straightforward to extend these
procedures and definitions to tensors of generic order $n$ (see
below) .\newline

\textbf{Remark 1 }The natural splitting of a tensor is
gauge-independent: it depends only on the physical frame chosen.
The projection technique gives us gauge-invariant quantities which
can have an operative meaning in our
physical frame; namely, they represent the objects of our measures.%
\newline

\textbf{Remark 2 } In $\Gamma$-adapted coordinates, a \textit{time
vector} $\mathbf{\bar{v}\in } \ \Theta _{p}$ is
characterized by the vanishing of its controvariant space components ($\bar{v%
}^{i}=0$); a \textit{space vector} $\mathbf{\tilde{v}\in } \
\Sigma _{p}$ by the vanishing of its covariant time component
($\tilde{v}_{0}=0$). As a generalization: (i) a given index of a
tensor $\mathbf{T}$ is called a \textit{time-index} if all the
tensor components of the type $T_{...}^{.i.}$ ($i \in [1,2,3]$)
vanish; (ii) a given index of a tensor $\mathbf{T}$ is called a
\textit{space-index} if all the tensor components of the type
$T_{.0.}^{...}$ vanish. For a \textit{time tensor}, i.e. for a
tensor belonging to $\Theta_p \ \otimes...\Theta_p$, property (i)
holds for all its indices; for a \textit{space tensor}, i.e. a
tensor belonging to $\Sigma_p \ \otimes ... \Sigma_p$, property
(ii)
holds for all its indices.\\

\indent \setcounter{subsection}{7} \textbf{\thesubsection
\ }\label{ssec:c7}To formulate the physical equations relative to the frame $%
\Gamma $, we need the following differential operator
\begin{equation}
\tilde{\partial}_{\mu }\doteq \partial _{\mu }-\gamma _{\mu }\gamma ^{{0}%
}\partial _{{0}}  \label{eq:dertras}
\end{equation}
which is called \textit{transverse partial derivative}. It is a "space
vector" and (its definition) is gauge-invariant.\newline

It is easy to show that, for a generic scalar field $\varphi (x)$
we obtain:
\begin{equation}
P_{\Sigma }(\partial _{\mu }\varphi )=\tilde{\partial}_{\mu }\varphi
\label{eq:protrans}
\end{equation}
So $\tilde{\partial}_{\mu }$ defines the \textit{transverse
gradient,} i.e. the space projection of the local
gradient.\newline

The projection technique we have just outlined allows to calculate the
projections of the Christoffel symbols. It is remarkable that the total
space projections read
\begin{eqnarray}
P_{\Sigma \Sigma \Sigma }(\mu \nu ,\lambda ) &=&\frac{1}{2}\left( \tilde{%
\partial _{\mu }}\gamma _{\nu \lambda }+\tilde{\partial _{\nu }}\gamma
_{\lambda \mu }-\tilde{\partial _{\lambda }}\gamma _{\mu \nu }\right) \doteq
\widetilde{\Gamma}^{*}_{\mu \nu \lambda}  \nonumber \\
P_{\Sigma \Sigma \Sigma }\left\{ _{\mu \nu }^{\lambda }\right\} &=&%
\widetilde{\Gamma}^{*}_{\mu \nu \sigma}\gamma ^{\sigma \lambda } \doteq
\widetilde{\Gamma}^{* \ \lambda}_{\mu \nu}  \label{eq:prochris}
\end{eqnarray}
where the space metric tensor $\gamma _{\mu \nu}$ substitutes the metric tensor $%
g_{\mu \nu }$ and the transverse derivative substitutes the
''ordinary'' partial derivative.\newline

\indent \setcounter{subsection}{8} \textbf{\thesubsection \
}\label{ssec:c8} \newcounter{CiOtto}\setcounter{CiOtto}{8} The
differential features of the congruence $\Gamma $ are described by
the following tensors
\begin{eqnarray}
\widetilde{C}_\mu &=&\gamma ^{\nu }\nabla _{\nu }\gamma _{\mu }
\label{eq:cmu} \\
\widetilde{\Omega }_{\mu \nu } &=&\gamma _{0}\left[ \widetilde{\partial }%
_{\mu }\left( \frac{\gamma _{\nu }}{\gamma _{0}}\right) -\widetilde{\partial
}_{\nu }\left( \frac{\gamma _{\mu }}{\gamma _{0}}\right) \right]
\label{eq:vortex} \\
\widetilde{K}_{\mu \nu } &=&\gamma ^{{0}}\partial _{{0}}\gamma _{\mu \nu }
\label{eq:born}
\end{eqnarray}
$\widetilde{C }_\mu $ is the \textit{curvature vector}, $\widetilde{\Omega }%
_{\mu \nu }$ is the \textit{space vortex tensor}, which gives the
local angular velocity of the reference fluid, $\widetilde{K}_{\mu
\nu }$ is the \textit{Born space tensor}, which gives the
deformation rate of the reference fluid; when this tensor is null,
the frame is said to be rigid according to the definition of
rigidity given by Born\cite{born}. In a relativistic context the
classical concept of rigidity, which is dynamical in its origin,
since it is based on the presence of forces that are responsible
for rigidity, becomes meaningless. The Born definition of rigidity
is the natural generalization of the classical one. It depends on
the motion of the test particles of the congruence, hence it is a
kinematical constraint. According to Born, a body  moves rigidly
if the space distance $\sqrt{\gamma_{ij}dx^i dx^j}$ between
neighbouring points of the body, as measured in their successive
(locally inertial) rest frames, is constant in time\footnote{In
the simple case of translatory motion, a body moves rigidly if, at
every moment, it has a Lorentz contraction corresponding to its
observed instantaneous velocity, as measured by an inertial
observer.}. For the Born condition see
Rosen\cite{rosen}, Boyer\cite{boyer}, Pauli\cite{pauli}.\\

\textbf{Definitions} The following definitions\footnote{%
It is worthwhile to notice that, in the literature, there is not
common agreement about these definitions, see f.i.
Landau-Lifshits\cite{landau}.} are referred to the (geometry of)
physical frame $\Gamma$:

\begin{itemize}
\item  \textit{constant} - when there exists at least one adapted chart, in
which the components of the metric tensor are not depending on the
time coordinate: $\partial _{0}g_{\mu \nu }=0$

\item  \textit{time-orthogonal} - when there exist at least one adapted chart
in which $g_{0i}=0$; in this system the lines $x^{{0}}=var$ are
orthogonal to the 3-manifold $x^{{0}}=cost$

\item  \textit{static} -  when there exists at least one adapted chart in which $%
g_{0i}=0$ and $\partial _{0}g_{\mu \nu }=0$.

\item  \textit{stationary} when it is constant and non time-orthogonal
\end{itemize}

\textbf{Remark 3 }The condition of being time-orthogonal is a
property of the physical frames, and not of the coordinate system:
for a reference frame to be time-orthogonal it is necessary and
sufficient that the space vortex $\widetilde{\Omega}_{\mu\nu}$
tensor vanishes.\\

When the space vortex tensor is null, moreover, the fluid is said
to be irrotational;  if both the curvature vector and the space
vortex tensor are zero, the fluid is said irrotational and
geodesic. Furthermore, when the space vortex tensor is not null, a
global synchronization of the standard clocks in the frame is not
possible.\\

The irrotational, rigid and geodesic motion (of a frame) is
characterized by the condition $\nabla _{\mu }\gamma _{\nu }=0$:
this is the generalization, in a curved spacetime context, of the
translational uniform motion in flat spacetime.\newline

\indent \setcounter{subsection}{9} \textbf{\thesubsection \
}\label{ssec:c9}The natural splitting permits also to calculate
the geometrical features, in particular the Riemann curvature
tensor of the 3-space of the reference frame (see section
\ref{sec:curvature}, below). In particular this will enable us to
define the curvature of the space of the rotating disk. Of course,
the result depends on the definition of the concept of 'space of
the disk ', which will be given in next section.

\section{The 'relative space' of a rotating platform} \label{sec:relspace}

\indent \setcounter{subsection}{1} \textbf{\thesubsection \
}\label{ssec:r1}As we shall see later, a rotating platform defines
a non-time-orthogonal physical frame; then the usual 3+1 foliation
of the spacetime, is meaningless. However, using the projection
technique, we  are going to introduce a different splitting, by
which we shall define the 'relative space' of the disk.\\ Let $K$
be an inertial frame with an adapted set of coordinates $\left\{
x^{\mu }\right\}~=~\left(t,r,\theta ,z\right) $, with line element
given by

\begin{equation}
\mathrm{d}s^{2}=g_{\mu \nu }dx^{\mu }dx^{\nu }=c^{2}\mathrm{d}t^{2}-\mathrm{d%
}r^{2}-r^{2}\mathrm{d}\theta ^{2}-\mathrm{d}z^{2}\mathrm{\ .}
\label{eq:metricapiatta}
\end{equation}
In this frame let us consider the equations
\begin{equation}
\left\{
\begin{array}{rcl}
r & = & r_{0} \\
\theta  & = & \theta _{0}+\omega t \\
z & = & z_{0}
\end{array}
\right. \mathrm{\ .}  \label{discorotante}
\end{equation}
If $r_{0}\in [0,R]\,,$ these equations describe (the points of) a
cylinder with radius $R$ rotating with constant angular velocity
$\omega $ (as measured in $K$). When $z_{0}$ is the same for each
point of the system, we deal with a rotating disk, whose points
have cylindrical coordinates $r_0,\theta_0$, representing their
initial positions ($t=0$).\\

The world-lines of each point of the disk are time-like helixes
(whose pitch, depending  on $\omega $, is constant), wrapping
around the cylindrical surface  $r~=~r_{0}~=~const$, with $r \in
[0,R]$. These helixes fill, without intersecting, the whole
spacetime region defined by $r\leq R<c/\omega $;  they
constitute a time-like congruence which defines the rotating frame $K_{rot}$%
, at rest with respect to the disk.\footnote{%
\smallskip The constraint $R<c/\omega $ simply means that the velocity of
the points of the disk cannot reach the speed of light. }

Let us introduce the coordinate transformation
\begin{equation}
\left\{
\begin{array}{rcl}
x'^{0} & = & t^{\prime }=t \\
x'^{1} & = & r^{\prime }=r \\
x'^{2} & = & \theta ^{\prime }=\theta -\omega t \\
x'^{3} & = & z^{\prime }=z
\end{array}
\right. \mathrm{\ .}  \label{catt}
\end{equation}

The coordinate transformation $\left\{ x^{\mu }\right\}
\rightarrow \left\{ x'^{\mu }\right\}$ defined by (\ref{catt}) (
which is not   internal to $K$ because of the time-dependence of
$x'^2$) has a kinematical meaning, namely it defines the passage
from a chart adapted to the inertial frame $K$ to a chart adapted
to the rotating frame $K_{rot}$. In fact along the helixes of the
congruence we obtain
\begin{equation}
\left\{
\begin{array}{rclcrcl}
\mathrm{d}x^{0\prime }=c\mathrm{d}t^{\prime } & \neq & 0 &
\Longleftrightarrow & c\mathrm{d}t & \neq & 0 \\
\mathrm{d}x^{1\prime }=\mathrm{d}r^{\prime } & = & 0 & \Longleftrightarrow &
\mathrm{d}r & = & 0 \\
\mathrm{d}x^{2\prime }=\mathrm{d}\theta ^{\prime } & = & 0 &
\Longleftrightarrow & \mathrm{d}\theta & = & \omega \mathrm{d}t \\
\mathrm{d}x^{3\prime }=\mathrm{d}z^{\prime } & = & 0 & \Longleftrightarrow &
\mathrm{d}z & = & 0
\end{array}
\right. \mathrm{\ ,}  \label{eli}
\end{equation}
which means that the lines $x'^{0}=var$  are exactly the
lines of the congruence defining $K_{rot}$.\\

In the chart $\left\{ x'^{\mu}\right\}$ the metric tensor is
written in the form\footnote{For the sake of simplicity, we
substitute $r'=r$, from (\ref{catt}) $_{\mathit{II}}$.}:
\begin{equation}
g_{\mu \nu }^{\prime }=\left(
\begin{array}{cccc}
1-\frac{\omega ^{2}{r}^{2}}{c^{2}} & 0 & -\frac{\omega {r}^{2}}{c} & 0 \\
0 & -1 & 0 & 0 \\
-\frac{\omega {r}^{2}}{c} & 0 & -{r}^{2} & 0 \\
0 & 0 & 0 & -1
\end{array}
\right)  \label{born}
\end{equation}
This is the so called Born metric, and in the classic textbooks
(f.i. \cite{moller},\cite{landau})it is commonly presented as the
spacetime metric in the rotating frame
of the disk.\\

\textbf{Remark 1 } In the chart $\left\{ x'^{\mu \ }\right\}
$\footnote{Born chart.} the time  $t'$ is equal to the coordinate
time $t$ of the inertial frame $K$. In this way, we label  each
event $P$ in $K_{rot}$ using the time of a clock at rest in $K$,
whose world-line (a straight line parallel to the time axis)
intersects $P$, and not by means of clock at rest on the disk. We
identify the coordinate time $t'$ with the time $t$ of the
observations of the events as measured by a clock in $K$. As
pointed out by Tangherlini\cite{tangherlini} and Gr\o
n\cite{gron3}, the transformation (\ref{catt}) has a Galilean
character, and this is due to the peculiarity of angular velocity
which, contrary to translational velocity, has an absolute value,
that can be locally measured.\\
The parameterization of the rotating frame $K_{rot}$ by the
coordinate $t$ of the inertial frame $K$ is the only way to
synchronize the clocks on the platform, whose proper times cannot
be synchronized by the Einstein's convention, because of the non
time-orthogonality of $K_{rot}$ (see eq. (\ref{eq:omega12}) below).\\

\textbf{Remark 2} Of course, if we do not care of global
synchronization on the disk, a different choice of the chart can
be made, which has a direct operational meaning for an observer on
the disk, by substituting the coordinate time $t$ with the proper
time of the clocks at rest in $K_{rot}$. Then we get the Post
chart, and the
generalized Born metric\cite{post}.\\

\indent \setcounter{subsection}{2} \textbf{\thesubsection \ }
 \label{ssec:r2} \newcounter{ErreDue} \setcounter{ErreDue}{2}
  Given the metric tensor in terms of coordinates adapted to the
rotating frame, we can compute the covariant components of the
vector field \mbox{\boldmath $\gamma$}$(x)$ of the congruence
$\Gamma$, using Equations (\ref {eq:gammas}):
\begin{equation}
\left\{
\begin{array}{rcl}
\gamma _{0}^{\prime } & = & \gamma ^{-1} \\
\gamma _{1}^{\prime } & = & 0 \\
\gamma _{2}^{\prime } & = & -\gamma \frac{\omega {r}^{2}}{c} \\
\gamma _{3}^{\prime } & = & 0
\end{array}
\right. \mathrm{\ \quad .}  \label{gamma-born-covar}
\end{equation}
where $\gamma \doteq 1/\sqrt{1-\frac{\omega ^{2}{r}^{2}}{c^{2}}}$
is the Lorentz factor of a point of the disk whose distance from
the rotation axis is $r^{\prime }=r\leq R$. As a consequence, in
the Born chart:\\

(i) the components of the space metric tensor are:
\begin{equation}
\gamma _{ij}^{\prime }=g_{ij}^{\prime }-\gamma _{i}^{\prime }\gamma
_{j}^{\prime }=\left(
\begin{array}{ccc}
-1 & 0 & 0 \\
0 & -\gamma ^{2}{r}^{2} & 0 \\
0 & 0 & -1
\end{array}
\right) \mathrm{\ \quad .}  \label{tms}
\end{equation}

(ii) the space vortex tensor $\tilde{\Omega}'_{\mu \nu }$ has the
following non-zero component(see Appendix, eq.
(\ref{eq:omega12all})):

\begin{equation}
\widetilde{\Omega^{\prime}}_{12}= -\gamma^3 \frac{2 \omega r}{c}\neq 0
\label{eq:omega12}
\end{equation}
which shows that the rotating frame $K_{rot}$ is not time-orthogonal.\\
For a rotating disk, it is easy to verify that the Born space
tensor vanishes:

\begin{equation}
\widetilde{K^{\prime}}_{ij}\doteq \gamma
^{^{\prime}0}\partial^{\prime}_{0}\gamma^{\prime}_{ij}=0
\label{eq:kappaij}
\end{equation}
since the space metric (\ref{tms}) does not depend on the time
coordinate\footnote{ In fact, in the chart $\{x'^{ \mu} \}$, the
steady motion condition ($\partial'_0 \equiv 0$) implies that the
metric tensor does not depends on $x'^{0}$; as a consequence we
obtain $ \tilde{\partial}'_\mu=\partial'_\mu$.}. Hence the
rotating frame $K_{rot}$ is rigid, in the sense of Born rigidity
(section \ref{sec:cattaneo}.\theCiOtto ).\\

\indent \setcounter{subsection}{3} \textbf{\thesubsection \
}\label{ssec:spazquoz} \newcounter{SpazQuoz}
\setcounter{SpazQuoz}{3} The concept of physical space is
\textit{locally} defined by the space platform $\Sigma _{p}$
normal to the time direction identified by the vector
\mbox{\boldmath $\gamma$} in the point $p$.\\

As we show in the Appendix\footnote{ See in particular eqs.
(\ref{eq:Ktr}), (\ref{eq:Kthetar}).}, the congruence $\Gamma$ of
time-like helixes, wrapping around the cylindrical hypersurfaces
$\sigma _{r}\,$ ($r=cost\in ]0,R]$), defines a Killing field not
in \textit{\ }$\mathcal{M}^{4}$, but on the submanifolds \textit{%
\ }$\sigma _{r}\;\subset \mathcal{M}^{4}$. We point out the
following interesting consequence:\textbf{\ }\textit{the splitting
}$T_{p}=\Theta _{p}\oplus \Sigma _{p}$\textit{\ and the space
metric tensor }$\gamma'_{ij}(p)$ are invariant along the lines of
$\Gamma$. It is then possible to define a one-parameter group of
diffeomorphisms with
respect to which both  the splitting $%
T_{p}=\Theta _{p}\oplus \Sigma _{p}$ and the space metric tensor
$\gamma'_{ij}(p)$ are invariant. The lines of $\Gamma$ constitute
the
trajectories of this "space~$\oplus$~time isometry".\\

This important property suggests a procedure to define an
\textit{extended} 3-space,  which we shall call \textit{`relative
space'}: it will be recognized as the only space having an actual
physical meaning  from an operational point of view, and it will
be identified as the 'physical space of the
rotating platform'.\\

\textbf{Definition. } Each element of the relative space is an
equivalence class of \textit{points and of space platforms}, which
verify this equivalence relation:\newline

RE\label{relequiv}:\textit{\ \noindent `` Two points (two space
platforms) are equivalent if they belong to the same line of the
congruence ''}.\newline

That is, the \textit{relative space} is the "quotient space" of
the world tube of the disk, with respect to the equivalence
relation RE, among points
and space platforms belonging to the lines of the congruence $\Gamma$.%
\newline

%\verb|begin addition| \\

This definition simply means that the relative space is the
manifold whose ''points'' \textit{are} the lines of the congruence
(see f.i Wahlquist-Estabrook\cite{waest}, Wahlquist\cite{wahl},
Rizzi-Tartaglia\cite {rizzi2},\cite{rizzi1} and
Norton\cite{norton}). However, our definition emphasizes the role
of the space platforms, which is often neglected: the reference
frame defined (as above) by the relative space coincides
\textit{everywhere} with the local rest frame of the rotating
disk.

We stress that it is not possible to describe the relative space
in terms of space-time foliation, i.e. in the form $x^{0}=const$,
where $x^{0}$ is an appropriate coordinate time, because the space
of the disk, as we saw before, is not time-orthogonal. Hence,
thinking of the space of the disk as a submanifold or a subspace
embedded in the space-time, as some author
claims \cite{weber1},\cite{berenda}, is misleading and meaningless\footnote{%
The best we can do, if we long for some kind of visualization, is
to think of the relative space as the union of the infinitesimal
space platforms, each of which is associated, by means of the
request of $M$-orthogonality, to one and only one line of the
congruence. }.\newline

\textbf{Remark 3 } The relative space is naturally endowed with a
3-dimensional (spatial) Riemannian structure, which is well
defined since it does not depend on the time variable along the
curves of the congruences, because of the vanishing of the Born
tensor. In particular, a ''spatial''
scalar product can be defined in the relative space. Consider two vectors $%
\mathbf{v}$ and $\mathbf{w}$ belonging to the tangent space {\normalsize $%
T_{p}=\Theta _{p}\oplus \Sigma _{p}$\textit{\ }}at a point $p \in
 \mathcal{M}^{4}$. According to the Cattaneo's procedure
outlined
in Section \ref{sec:cattaneo}, we can project these vectors to obtain two spatial vectors  $\mathbf{%
\widetilde{v}}$ and $\mathbf{\widetilde{w}}$, belonging to $\Sigma
_{p}$, now interpreted as the tangent space to the relative space
in $p$ (which  here is to be intended as the line of the
congruence passing through $p \in \mathcal{M}^{4}$). In the chart
$\left\{ x^{\prime }{}^{\mu }\right\} $ their covariant components
are
\begin{equation}
\widetilde{v}_{i}^{\prime } = \gamma _{ij}^{\prime }v^{\prime
}{}^{j}  \ \ \ \widetilde{w}_{i}^{\prime } = \gamma _{ij}^{\prime
}w^{\prime }{}^{j} \label{eq:vet_SpazRel}
\end{equation}
where $\gamma _{ij}^{\prime }$ is the space metric tensor defined
in eq. (\ref{tms}).  So the scalar product $(\ ,\ )_{RS}$ in the
(tangent bundle of the) relative space is defined by
\begin{equation}
(\mathbf{\widetilde{v}},\mathbf{\widetilde{w}})_{RS}\doteq
\gamma'^{ij}\widetilde{v}'_i \widetilde{w}'_j
\label{eq:scalar_prod_Spaz_Rel}
\end{equation}

%\verb|end addition|

\section{Lengths in the relative space} \label{sec:lengths}

\begin{figure}[top]
\begin{center}
\includegraphics[width=8cm,height=8cm]{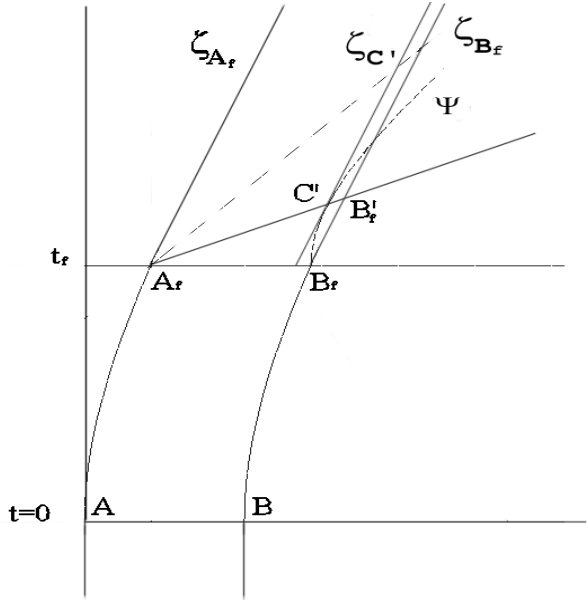}
\caption{\small The world-strip $(\zeta_{A_f},\zeta_{C'})$ of a
length $\lambda_0=AB=A_f B_f$, at rest on the rim, does not cover
entirely the world-strip $(\zeta_{A_f},\zeta_{B_f})$ of this
length, as measured in the inertial frame.} \label{fig:figura}
\end{center}
\end{figure}

We did not clarify yet, in operational terms, the identification
of the "relative space" with the physical space of the disk. In
order to do that, we start by analyzing what happens to "standard
lengths" when they undergo an acceleration process.\\
Let us consider the world-strip of an infinitesimal piece of the
rim of the disk, which is at rest until $t=0$ in the inertial
frame $K$ (see figure \ref{fig:figura}).

When $t=0$, the disk starts being accelerated in such a way that
all points of its rim have identical motion, as observed in $K$.
If $I=[0,t_f]$ is the interval  representing the period of time
during which acceleration acts, $\forall t \in I$ the acceleration
distribution of all points of the rim is the same, as observed in
the inertial frame $K$. From a pictorial point of view, this means
that the world lines of all points of the rim are congruent (i.e.
superimposable). During the acceleration period, the disk is not
Born-rigid\cite{gron1} although it appears rigid in $K$. This
means that, depending on the simultaneity criterium in the
inertial frame, the length of the infinitesimal piece of the rim
is always congruent with the starting segment $AB$; in particular,
when $t=t_f$, it is represented by the segment $A_f B_f$. On the
other hands, from the point of view of the local observer at rest
on the rim, whose world line passes through $A$ when $t=0$, the
simultaneity criterium is not defined by the family of straight
lines parallel to $AB$, but it varies at each instant, depending
on the velocity (in $K$) of the rim itself. Namely, when the
acceleration period finishes, the piece of the rim  is represented
by the segment $A_f B'_f$, in the local co-moving frame associated
with $A_f$. Let us put $A_f B_f=AB=\lambda_0$, where $\lambda_0$
is the wavelength of a monochromatic radiation emitted by a source
at rest in $K$\footnote{Since $AB$ is infinitesimal, the
monochromatic radiation must be chosen in such a way that
$\lambda_0$ is very small when compared with the length of the
circumference.}. The $M$-circumference of radius $\lambda_0$, with
center in $A_f$, whose equation is
\begin{equation}
\Psi \equiv \{P\in M^{2}:A_{f}P=\lambda _{0}\}
\label{eq:Mcircumference}
\end{equation}
%\Upsilon
can be built by  considering, in each reference frame, the
wavelength of the given radiation, emitted by a source at rest in
that frame. This $M$-circumference, which is a hyperbola in the
Minkowskian plane, intersects the segment $A_{f}B_{f}^{\prime }$
in $C^{\prime }$, and we obtain
\begin{equation}
A_{f}B_{f}^{\prime }=A_{f}C^{\prime }\gamma =\lambda _{0}\gamma
>\lambda _{0} \label{eq:Mcirc}
\end{equation}
This relation means that the world-strip
$(\zeta_{A_f},\zeta_{C'})$ of a length $\lambda_0$, at rest on the
rim, does not cover entirely the world-strip
$(\zeta_{A_f},\zeta_{B_f})$ of this length, as measured in $K$
(see figure \ref{fig:figura}). From a physical point of view,
equation (\ref{eq:Mcirc}) shows that each element of the periphery
of the disk, of proper length $\lambda_0$, is stretched during the
acceleration period. This a
purely kinematical result of our acceleration program.\\
However, this result remains valid if one takes into account the
interactions among the physical points of the rim (f.i. those
interactions which ensure rigidity in the phase of stationary
motion). In particular, during and after the acceleration period,
each point of the disk is subject to both radial and tangential
stresses; the former maintain each point on the circumference
$r=r_0$, while the latter ought to give zero resultant, because of
the axial symmetry: \textit{each point is pulled in the same way
by its near points, in both directions}. As a consequence, the
elongation of every element of the rim, due to tidal forces
experienced during the acceleration period, remains even when
acceleration finishes\footnote{Let us point out that a shortening
of the elements of the rim, due to Hooke's law, cannot be invoked,
since these elements have not free
endpoints.}.\\

\normalsize

\textbf{Remark } The arguments given treat the disk as a set of
non interacting particles. The only constraint is that every
particle must move along a circular trajectory, with a given law
of motion, according to a kinematical definition of rigidity in
$K$ . In our study, we decided to neglect the radial dilation and
deformation effects which are present in a real body when it is
set into rotation; they would cause an enlargement of the radius,
and a consequent enlargement of the circumference. Our purpose is
not the study of the elastic deformation of a real disk, but
working out the relativistic corrections to the "rigid" motion of
a rotating disk.\\

From the considerations above, it follows that the dilation which
is responsible for the Ehrenfest's paradox  has a pure kinematical
origin. The enlargement of the rod (assumed as a standard rod), in
the rest frame  at the end of acceleration phase, is due to the
change of the simultaneity criterium. In figure \ref{fig:figura}
this is represented by the change in the slope of the
infinitesimal space platforms which are associated, by means of
the request of $M$-orthogonality, to the lines of the congruence
$\Gamma$.

\subsection{Optical congruence and "S{\`{e}}vres meters"
congruence} \label{ssec:congruence}

As well known, Einstein had been stressing, since the early years
of the XX century, that the space geometry is determined by the
empirical properties of the bodies (thought as rigid ones), and it
can be explored using measuring rods, superimposable by means of
displacements (i.e. congruent).\\
Hence, a complete operational approach to a relativistic problem
cannot be done without the definition of the \textit{congruence},
intended as the measurement process of distances, by means of the
confrontation with rigid  rods (i.e. standard meters),  slowly
transported in  space. Indeed, the standard meter  is, in
principle, arbitrarily chosen\footnote{Every change in the choice
of the standard meter, and consequently in the choice of the
congruence, provokes a change of the geometry and, in particular,
a change of the metric. Hence the metric is not determined
univocally by the physical situation, but it depends on the
definition of the congruence which is adopted. So the concepts of
metric and congruence are not observation-dependent, but are
conventional, and only their union must be seen as physically
meaningful (Reichenbach in \cite{reichenbach2} and
\cite{reichenbach3}, with a reply by Einstein \cite{einstein3}).},
nevertheless it must have the following properties:

\begin{enumerate}
\item Two standard meters, which have the same chemical and physical
composition and which are superimposable at a given instant,
remain superimposable at any time, after an arbitrary sequence of
displacements.
\item In the class of inertial frames (local or global),  light
rays in vacuum propagate in straight lines with respect to the
standard meter.
\end{enumerate}

Condition (2), by which we can  substitute the light rays for the
rigid rods, allows us to assume (locally) the trajectory of a
light ray as a space geodesic\footnote{ The trajectories of light
rays, even if they are always geodesics of the space time, are
not, in general, geodesics of the three-dimensional space (in
particular they are not in general geodesics of the relative
space).}, and as unit of length the wavelength in vacuum of a
given spectral line, emitted by a source at rest. This congruence
is called \textit{optical congruence} and it the defines the
\textit{optical space}\footnote{Reichenbach\cite{reichenbach2}
uses the terms \textit{light space} and \textit{light congruence}
to refer to the \textit{optical space} and the \textit{optical
congruence}, respectively.}.\\

It is possible to adopt a different choice of the standard meter,
for instance the one that we use normally to perform space
measurements in ordinary scale: we shall call it the "congruence
of S\`{e}vres meters". In this congruence: (i) the standard rod,
intended as rod of a given length, remains the same when slowly
transported; (ii) in different metric conditions, that is in
presence of a gravitational or inertial field, each rod maintains
the same length that it would have in absence of acceleration.\\

\textbf{Remark 1 } It is interesting to notice that convention
(ii), which was actually used by Einstein,  treats in a different
way the \textit{metric} field (gravitational and inertial) and the
three \textit{physical} fields (electromagnetic, strong and
weak)\footnote{Using the words of Reichenbach \cite{reichenbach},
the gravitational field is \textit{universal}, i.e. it acts
equally on each test body, according to the equivalence principle;
the physical fields are \textit{differential}, in the sense that
they act differently on test bodies with different
chemical-physical composition. It is well known that this is the
\textit{physical}  reason for which the GRT geometrizes the
gravitational field, which becomes a metrical field in the
spacetime, but does not geometrize the physical fields, which act
in the spacetime.}. Locally the meter of S\`{e}vres is subject
only to the physical fields, which deform it; however this
deformation can be corrected for, according to physical laws. As a
consequence  "corrected standard meters of S\`{e}vres" are
obtained. So the geometry of
the space is actually explored by the "meters of S\`{e}vres", in the physical frame where they are at rest.\\

\textbf{Remark 2 } It is essential to stress  that the congruence
of the S\`{e}vres meters defined above uses rods with free
endpoints. In the context of the  measurements performed on the
rotating disk, it does not appear correct using this convention
for a meter which is, actually, part of the circumference, as
Tartaglia\cite{tartaglia} claims. The meter cannot be transported
in  space without cutting its endpoints: because of the cuts, the
tangential force acting upon the endpoints, which compensated each
other before the cuts for symmetry reasons, provoke the
shortening of the rod, according to the Hooke's law.\\

\textbf{Remark 3 } We want to stress that the use of the
\textit{optical congruence} is meaningful only in any
\textit{local} Minkowskian (tangent) frame, whose space geometry
is actually the geometry of an optical space. However, the
\textit{global} geometry of the relative space, induced by the
geometry of the space-time and by the congruence of the world
lines which defines the space platform, \textit{is not}  an
optical space, neither it can be reduced to an optical space using
a conformal transformation (unlike the case of a static space
\cite{Abramowicz}). Hence the relative space, endowed with the
metric tensor $\gamma'_{ij}$\thinspace , is an optical space only
locally, but not globally.

\section{Curvature of the relative space}\label{sec:curvature}
After having introduced and described the relative space, now we
can characterize its geometry, that is its metric and curvature,
using the techniques we introduced in section  \ref{sec:cattaneo}.\\
The measurements of lengths in the relative space, are done using
the space metric tensor $\gamma^{\prime}_{ij}$:

\begin{equation}
\gamma _{ij}^{\prime }=\left(
\begin{array}{ccc}
-1 & 0 & 0 \\
0 & -\gamma ^{2}{r}^{2} & 0 \\
0 & 0 & -1
\end{array}
\right)   \label{eq:metricagamma}
\end{equation}

The rotating observer can perform measurements of both space and
time on the platform. Measurements of time are performed, by the
observer, using its own standard clocks, on which he reads the
proper time. Measurements of space are performed without care of
time, since the metric (\ref{eq:metricagamma}) does not depend on
time\footnote{This is a consequence of the stationarity of the
rotating frame.}.\\  Using the Cattaneo's projection
technique\cite{cattaneo}, the curvature tensor of the relative
space can be defined\footnote{ For the sake of simplicity, in this
section, we remove the primed letters: however we shall always
refer to the coordinates$\{x'^{\mu}\}$ adapted to the rotating
frame.}: \small
\begin{eqnarray}
\widetilde{R}_{\mu \nu \rho \sigma }^{*} &\doteq &\frac{1}{4}\left(
\widetilde{\partial }_{\rho \mu }\gamma _{\nu \sigma }-\widetilde{\partial }%
_{\sigma \mu }\gamma _{\nu \rho }+\widetilde{\partial }_{\sigma \nu }\gamma
_{\mu \rho }-\widetilde{\partial }_{\rho \nu }\gamma _{\mu \sigma }\right) +%
\frac{1}{4}\left( \widetilde{\partial }_{\mu \rho }\gamma _{\nu \sigma
}-\right.  \nonumber \\
&&\left. \widetilde{\partial }_{\nu \rho }\gamma _{\mu \sigma }+\widetilde{%
\partial }_{\nu \sigma }\gamma _{\mu \rho }-\widetilde{\partial }_{\mu
\sigma }\gamma _{\nu \rho }\right) +\gamma ^{\alpha \beta }\left[ \widetilde{%
\Gamma }_{\sigma \nu ,\alpha }^{*}\widetilde{\Gamma }_{\rho \mu ,\beta }^{*}-%
\widetilde{\Gamma }_{\rho \nu ,\alpha }^{*}\widetilde{\Gamma }_{\sigma \mu
,\beta }^{*}\right]  \label{eq:curvspaz}
\end{eqnarray}
\normalsize where the space Christoffel symbols are defined in eq.
(\ref{eq:prochris}), and the transverse derivatives, defined by
eq. (\ref{eq:dertras}), are the same as the ordinary partial
derivatives because of the stationarity of the motion(see
subsection \ref{sec:relspace}.\theErreDue ). Since it has all
space indices (see Remark 2, subsection
\ref{sec:cattaneo}.\theCiSei ), the curvature tensor
(\ref{eq:curvspaz}) is a space tensor. Then the curvature tensor
which is adequate to describe the geometry of the relative space
of the disk is the space part $\widetilde{R}_{ijkl}^{*}$ of the
tensor (\ref{eq:curvspaz}). We have to compute the curvature
tensor of a three-dimensional riemannian manifold, whose metric
tensor is given in eq. (\ref{eq:metricagamma}). It is useful to
invert the signature, since we are dealing with the spatial
geometry:
\begin{equation}
\gamma _{ij}=\left(
\begin{array}{ccc}
1 & 0 & 0 \\
0 & \gamma ^{2}r^{2} & 0 \\
0 & 0 & 1
\end{array}
\right) \label{eq:invseg}
\end{equation}
The only non zero components of the curvature tensor of space of
the disk are
\begin{equation}
\widetilde{R}_{1212}^{*}=-3\frac{\left( \frac{\omega r}{c}\right) ^{2}}{%
\left( 1-\left( \frac{\omega r}{c}\right) ^{2}\right) ^{3}}=-3\,\beta
^{2}\,\gamma ^{6}  \label{riemann}
\end{equation}
and those which are obtained by the symmetries of the indices.\\
The non null components of the space projection of the Ricci
tensor, are:
\begin{equation}
\begin{array}{rcl}
\widetilde{R}_{11}^{*} & = & -3\frac{\omega ^{2}}{c^{2}}\,\gamma ^{4} \\
\widetilde{R}_{22}^{*} & = & -3\,\beta ^{2}\,\gamma ^{6}
\end{array}
\label{ricci}
\end{equation}
Finally the curvature scalar is:
\begin{equation}
\widetilde{R}^{*}=-6\,\frac{\omega ^{2}}{c^{2}}\,\gamma ^{4}  \label{scalar}
\end{equation}

We see that the geometry of the relative space of the disk has a
space curvature which is not zero, hence its geometry is not
Euclidean. Doing so, we succeeded in verifying Einstein's
intuition on the curvature of the rotating disk, after having
defined the geometrical context in which the curvature is
computed, i.e. the relative space.\\
It is interesting to notice that the results we obtained are in
agreement with the ones obtained by
Berenda\cite{berenda},Arzeli\`{e}s\cite{arzelies}, M\o
ller\cite{moller},Gr\o n\cite{gron1} who, nevertheless, do not
define explicitly the geometrical context (see subsection
\ref{sec:relspace}.\theSpazQuoz ). Furthermore, their calculations
do not rely upon the use of a splitting technique, like ours: they
just computed the components of the curvature tensor of the
\textit{space-time} which have all spatial indices ($R_{ijkl}$),
and they referred to it as the \textit{space} curvature tensor.
This is not correct, since a splitting procedure is needed to
obtain quantities that have a true physical meaning, i.e. which
are gauge invariant and, hence, observable. Nevertheless their
results are equal to ours, and this is due to the fact that for
the rotating disk in uniform motion, the physical frame $\Gamma$
is stationary, hence $\partial_0 \equiv 0$ and then
$\tilde{\partial}_\mu=\partial_\mu$ everywhere. In this case
$R_{ijkl}=\widetilde{R}_{ijkl}^{*}$. However, things are different
for those physical reference frames which lack symmetries, such as
the axis-symmetry and the
stationarity of the rotating disk.\\

\textbf{Remark } The complete space projection of the curvature
tensor of the space-time is \cite{cattaneo}:
\begin{equation}
P_{\Sigma \Sigma \Sigma \Sigma}\left(R_{\mu \nu \sigma \rho} \right)=%
\widetilde{R}_{\mu \nu \sigma \rho}^{*}-\frac{1}{4}\left(\tilde{\Omega}%
_{\sigma \mu} \tilde{\Omega}_{\rho \nu}-\tilde{\Omega}_{\rho \mu}\tilde{%
\Omega}_{\sigma \nu} \right)-\frac{1}{2} \tilde{\Omega}_{\mu \nu} \tilde{%
\Omega}_{\sigma \rho} =0  \label{eq:rie_space}
\end{equation}
where we have taken into account the fact that the curvature
tensor of the space-time $R_{\mu\nu \sigma \rho}$ is
null\footnote{The curvature we computed above
(eqs.(\ref{riemann})-(\ref{scalar})) refers to the
three-dimensional physical space of the disk, formally defined by
the relative space. It is superfluous to say that the curvature
tensor of the Minkowskian space-time is null, in each chart used
to compute it, and in each frame to which the chart is
adapted\cite{berenda}.}. This shows that the  space components
$\widetilde{ R}_{\mu \nu \sigma \rho }^{*}$ are completely defined
by the terms containing the space vortex tensor, which is related
to rotation: hence the non Euclidean nature of the space of the
disk depends only on the rotation of the frame.

\section{Measurements of lengths in the relative space} \label{sec:measure}

Now, the Ehrenfest's paradox can be solved, in a natural way, in
the
context of special relativity.\\
The measurements of lengths along the rim of the disk are
determined by the space metric tensor $\gamma'_{ij}$, given (in
the inverted signature) by Eq. \ref{eq:invseg}. As a consequence,
the length of an infinitesimal segment on the rim of the
circumference is
\begin{equation}
dl^{\prime}_{\Sigma}=\left(\sqrt{\gamma_{ij}'(\omega,r')dx'^i
dx'^j} \right)_{r'=R,z=cost}=\gamma(\omega,R)R d\theta'
\label{eq:dlsigma}
\end{equation}
From (\ref{catt})$_{III}$ it follows that at fixed coordinate time
of the inertial frame $K$, $d\theta'=d\theta$. Consequently, the
angle all around the periphery of the disk,
measured on it, is equal to $2\pi$: $\theta' \in [0,2\pi]$.\\
Hence, the measurement of the circumference on the rim of the
disk, performed by the rotating observer, turns out to be :
\begin{equation}
l^{\prime}_{\Sigma}= 2 \pi R \gamma  \label{eq:lsigma}
\end{equation}
where $\gamma = 1/\sqrt{1-\frac{\omega ^{2}{r}^{2}}{c^{2}}}$. This
is in agreement with the fact that the space geometry of the disk
is not Euclidean\footnote{Rizzi-Tartaglia\cite{rizzi1} and
Cantoni\cite{cantoni}  computed the length of the circumference
and reached the same result. The main difference between their and
our approach is that, in spacetime, the "circumference" considered
by these authors is not a closed curve, but an open, space-like
curve (the "helix of simultaneity"), because of the time lag due
to rotation. We believe that our approach is more adequate,
because lengths are measured without caring of time; on the
contrary in these authors' works, time enters explicitly in space
measurements, in spite of the impossibility of synchronization on
the platform.}.\\
For the observer in the inertial frame the length given in eq.
(\ref{eq:dlsigma}) appears contracted by the standard factor
$\gamma^{-1}$:
\begin{equation}
dl=\gamma^{-1}dl^{\prime}_{\Sigma}  \label{eq:dlI}
\end{equation}
Since
\begin{equation}
dl=\gamma^{-1}\gamma R d\theta^{\prime}= Rd\theta^{\prime}  \label{eq:dlI2}
\end{equation}
we obtain that, in correspondence of the measure of the
circumference given in eq. (\ref{eq:lsigma}), performed by the
rotating observer, the inertial observer measures a length $2\pi
R$, as expected, since the space of the inertial frame $K$
is Euclidean.\\

In conclusion, a proper definition of the natural space in which
the measurements are performed by the observer on the disk, leads
to the solution of the Ehrenfest's paradox. Moreover, it has been
showed that the space of the disk is not Euclidean.\\
We stress that everything has been done using the special
relativistic kinematics, without any \textit{ad hoc} dynamical or
kinematical hypotheses.

\section{Discussion and conclusions} \label{sec:conclusion}

The solution of the  Ehrenfest's paradox, that we outlined in this
paper, is strongly dependent both on a proper definition of the
physical space of the disk and a proper choice of the congruence
adopted to perform the measurement in such a space. Hence, the
introduction of the \textit{relative space} and the (local) use of
the \textit{optical congruence} are the bases of our results,
together with the use of a splitting procedure which allows a
correct (and intuitive) geometrical description of the concepts we
introduced.

Even if we believe that the study of the rotating disk, in the
context of relativistic dynamics, is certainly interesting, we
showed that a dynamical approach is not necessary to solve the
paradox, as some authors claimed in the past, like the mentioned
above Lorentz\cite{lorentz}, Eddington\cite{eddington},
Clark\cite{clark}, Cavalleri\cite{cavalleri}, Brotas\cite{brotas},
McCrea\cite{mccrea}.

Different assumptions both on the physical space and the physical
congruence are the main differences between our paper and the
works of those authors who tried to solve the paradox using
relativistic kinematics. Among them, Berenda\cite{berenda},
Arzeli\`{e}s\cite{arzelies},
Landau-Lifschits\cite{landau}, M\o ller\cite{moller}, Rosen\cite{rosen}, Gr%
\o n\cite{gron1},\cite{gron2}, Wahlquist-Estabrook\cite{waest}
obtained the same results for the metric tensor of the space of
the disk as those we obtained in the previous section, even if all
these works lack a clear and consistent definition of the space of
the disk itself\footnote{However, we must point out that
Wahlquist-Estabrook\cite{waest}, using a $3+1$ tetrad method which
is very similar to the Cattaneo's technique, worked out the
intrinsic geometry of a rotating and accelerating platform,
generalizing the results of the authors quoted above.}.

Klauber's\cite{klauber},\cite{klauber2} solution of the paradox is
ultimately a negation of it, since he claims that no Lorentz
contraction exists for rotating systems, and concludes that the
space of the disk is flat. According to him, Minkowskian tangent
frames are locally equivalent to accelerated frames only when the
latter are time-orthogonal. Such a local equivalence is refused
for a rotating platform, which is not a time-orthogonal frame. As
a consequence, he claims that, in the case of rotation, a new
theory, alternative to SRT, is needed; and therefore he proposes a
''New Theory of Rotating Frames''. We point out that: (i)
Klauber's assumptions can be maintained if and only if a
non-Einstein synchronization procedure is adopted on the platform
(see f.i. Weber \cite{weber}); (ii) any theory alternative to SRT
is not needed at all, in fact  we have just shown how the paradox
can be solved  in a strictly special relativistic context.

Tartaglia \cite{tartaglia} tries to solve the paradox assuming
that, in the case of rotations, the relativity of simultaneity
does not enter the space measurement processes of the observer
which is at rest on the platform. Therefore, he concludes that the
length of the circumference of the platform cannot be altered by
rotation, and the space of the disk remains Euclidean: ''in the
rotating observer's view nothing happened to the geometry of the
platform as a consequence of the rotation''. To give an
operational definition of the space of the disk, compatible with
this assumption, he introduces what we call a ''congruence of
S\`{e}vres meters'' on the platform, whose lengths are not
affected by rotation. However, it is essential to stress that the
''S\`{e}vres meters'' used by Tartaglia, actually, are parts of
the circumference; on the other hands, a standard S\`{e}vres meter
is a rod\textit{\ with free endpoints}. The ''S\`{e}vres meter''
used by Tartaglia cannot be transported in space without cutting
its endpoints: because of the cuts, the tangential forces acting
upon the endpoints, which compensated each other before the cuts
for symmetry reasons, provoke the shortening of the rod, according
to the Hooke's law (see subsection \ref{ssec:congruence}, Remark
2). Only after the cuts, the rod has become a standard S\`{e}vres
meter, whose length is not affected by rotation, according to the
hypothesis of locality\footnote{This expression, which is one the
most important axioms of SRT, states the local equivalence of an
accelerated observer with a momentarily comoving inertial
observer, provided that standard rods and clocks are used (see
Mashhoon\cite{mashh}).}. Tartaglia's claim ''no Lorentz
contraction'' in the case of rotation rests upon his assumption
that ''no synchronization is needed'' for space measurements on
the platform. On the contrary, our approach shows that
\textit{synchronization is essential, and it is incorporated in
the very definition of the relative space, through the equivalence
relation RE given in subsection \ref{sec:relspace}.\theSpazQuoz}.
In particular the criterium of simultaneity, related to
synchronization, is geometrically interpreted as the tilt of the
local space platforms, belonging to any world line of the
congruence $\Gamma $. \textit{When synchronization is neglected,
rotation itself is neglected}: as a consequence, it
is not surprising that no Lorentz contraction is found.\\

We believe that the misunderstandings in the theoretical and
operational treatments quoted above rely on the lack of clear and
self consistent definitions of the fundamental concepts used.

From an operational point of view, our approach rests upon a
precise choice of the standard rods used by the observer on the
platform, which are the ''standard optical rods'' used in
relativity. From a theoretical point of view, our approach is
based on a clear and consistent definition, in a strictly SRT
context, of the concept of the ''space of the disk'', which is
formally identified with the ''relative space'', through
Cattaneo's natural splitting and a suitable equivalence relation.

The geometry of the space of the disk, which follows from our
assumptions, turns out to be non Euclidean, according to the early
Einstein's intuition. Moreover, the relativistic kinematics
reveals to be self consistent, and able to solve the Ehrenfest's
paradox  without any need of dynamical considerations.

It is interesting to stress that the space metric which we
obtained coincides with the one found in classic textbooks of
relativity, \textit{in spite of a non trivial shift of the
context}.

We want to point out that the calculations of the curvature of the
space of the disk, which are present in the literature, sometimes
without a definite splitting procedure, are misleading, even when
the final results are formally correct. In fact these calculations
would lead to incorrect results for those physical reference
frames which lack symmetries, such as the axis-symmetry and the
stationarity of the rotating disk.

In conclusion, we showed that the SRT, even when applied to
rotating platforms, is self-consistent and does not raise
paradoxes, provided that proper definitions of geometrical and
kinematical entities are adopted. On the contrary, non standard
\textit{ad hoc} hypotheses about Lorentz contraction and
relativity of simultaneity, inconsistent with the axioms of the
SRT, arise when some ambiguously defined entities are adopted.\\
\\
\\

\textbf{Acknowledgements } We are deeply in debt to Dr. Alessio
Serafini, who studied the problem of the rotating disk in his
graduation thesis, under the supervision of one of us (G.R.).

\appendix

\section{Appendix}

\textbf{Notation} In this appendix, for the sake of simplicity, we
do not use primed letters: however we always shall refer to the
set of coordinates $\{x'^{\mu}\}$ which are adapted to the
rotating frame
(\ref{catt}).\\

The metric tensor expressed in coordinates adapted to the rotating
frame is
\begin{equation}
g_{\mu \nu }=\left(
\begin{array}{cccc}
1-\frac{\omega ^{2}{r}^{2}}{c^{2}} & 0 & -\frac{\omega {r}^{2}}{c} & 0 \\
0 & -1 & 0 & 0 \\
-\frac{\omega {r}^{2}}{c} & 0 & -{r}^{2} & 0 \\
0 & 0 & 0 & -1
\end{array}
\right)  \label{born1}
\end{equation}
The inverse to the metric tensor is:
\begin{equation}
g^{\mu \nu }=\left(
\begin{array}{cccc}
1 & 0 & -\frac{\omega }{c} & 0 \\
0 & -1 & 0 & 0 \\
-\frac{\omega }{c} & 0 & -\frac{1-\frac{\omega ^{2}{r}^{2}}{c^{2}}}{r^{2}} &
0 \\
0 & 0 & 0 & -1
\end{array}
\right)  \label{bornup}
\end{equation}

The non zero Christoffel symbols turn to be
\begin{eqnarray}
\Gamma _{00}^{\ \ 1} &=&-\frac{\omega ^{2}}{c^{2}}r  \label{christoffel} \\
\Gamma _{01}^{\ \ 2 } &=&\frac{\omega }{cr}  \nonumber \\
\Gamma _{02 }^{\ \ 1} &=&-\frac{\omega }{c}r  \nonumber \\
\Gamma _{12 }^{\ \ 2 } &=&\frac{1}{r}  \nonumber \\
\Gamma _{22 }^{\ \ 1} &=&-r  \nonumber
\end{eqnarray}
The covariant components of the congruence vectors are:
\begin{eqnarray}
\gamma _{0} &=&\sqrt{1-\frac{\omega ^{2}r^{2}}{c^{2}}}
\label{gammacovarianti} \\
\gamma _{1} &=&0  \nonumber \\
\gamma _{2 } &=&-\left( \sqrt{1-\frac{\omega
^{2}r^{2}}{c^{2}}}\right)
^{-1}\frac{\omega r^{2}}{c}  \nonumber \\
\gamma _{3} &=&0  \nonumber
\end{eqnarray}
while the controvariant ones are:
\begin{eqnarray}
\gamma ^{0} &=&\left( \sqrt{1-\frac{\omega
^{2}r^{2}}{c^{2}}}\right) ^{-1}
\label{gammacontrovarianti} \\
\gamma ^{1} &=&0  \nonumber \\
\gamma ^{2 } &=&0  \nonumber \\
\gamma ^{3} &=&0  \nonumber
\end{eqnarray}
The non zero components of the space vortex tensor are:
\begin{equation}
\widetilde{\Omega}_{12}=\gamma_0 \tilde{\partial_1}\left(\frac{\gamma_2}{%
\gamma_0} \right) =\gamma_0 \partial_1 \left(\frac{\gamma_2}{\gamma_0} \right)=-\frac{%
1}{c}\frac{2 \omega r}{(1-\frac{\omega^2 r^2 }{c^2})^{3/2}} = -\gamma^3\frac{%
2\omega r}{c} \label{eq:omega12all}
\end{equation}

The covariant components of the Killing tensor  of the given
congruence are
\begin{equation}
K_{\mu \nu}\equiv \gamma _{\mu ;\nu }+\gamma _{\nu ;\mu
}=\frac{\partial \gamma
_{\mu }}{\partial x^{\nu }}+\frac{\partial \gamma _{\nu }}{\partial x^{\mu }}%
-2\Gamma _{\mu \nu }^{\ \ \alpha }\gamma _{\alpha }
\label{eq:killinall}
\end{equation}
Taking into account (\ref{gammacovarianti}) and
(\ref{christoffel}), we obtain that the only non null components
in $\mathcal{M}^{4}$ turn to be:
\begin{equation}
K_{01}\equiv \gamma _{0;1}+\gamma _{1;0}=\frac{\partial \gamma _{0}}{%
\partial r}-2\Gamma _{01}^{\ \ 2 }\gamma _{2 }=\frac{\partial
\gamma _{0}}{\partial r}-g^{2 \alpha }\frac{\partial g_{0\alpha }}{%
\partial r}  \label{eq:terre}
\end{equation}
\begin{equation}
K_{2 1}\equiv \gamma _{2 ;1}+\gamma _{1;2 }=\frac{\partial \gamma
_{2 }}{\partial r}-2\Gamma _{2 1}^{\ \ 2 }\gamma
_{2 }=\frac{\partial \gamma _{2 }}{\partial r}-g^{2 \alpha }%
\frac{\partial g_{2 \alpha }}{\partial r} \label{eq:thetaerre}
\end{equation}

Then the components $K_{01}$, $K_{21}$ depend solely on the
partial derivatives with respect to $r$ of some functions of $r$.
If we evaluate these components in $\mathcal{M}^{4}$, we obtain a
non zero result, while if we evaluate the same components on the
cylindrical hypersurface $\sigma _{r} \equiv \{ r=const\;(>0)\}$,
they result identically zero. Summing up, we get:
\begin{equation}
K_{01}\equiv \gamma _{0;1}+\gamma _{1;0}=\left\{
\begin{array}{c}
\frac{\omega ^{2}r}{c^{2}}\left( \sqrt{1-\frac{\omega ^{2}r^{2}}{c^{2}}}%
\right) ^{-1}\;\neq 0\;\mathit{\ in \mathit{\ }}\mathcal{M}^{4}\mathit{\ }
\\
0\;\mathit{on \ the \ submanifold \ }\sigma _{r}\;(r=const)
\end{array}
\right.  \label{eq:Ktr}
\end{equation}
\begin{equation}
K_{2 1}\equiv \gamma _{2 ;1}+\gamma _{1;2 }=\left\{
\begin{array}{c}
-\frac{\omega ^{3}r^{3}}{c^{3}}\left( \sqrt{1-\frac{\omega ^{2}r^{2}}{c^{2}}}%
\right) ^{-3}\neq 0\ \;\;\mathit{in \mathit{\ }}\mathcal{M}^{4} \\
0\;\mathit{on \ the \ submanifold \ }\sigma _{r}\;(r=const)
\end{array}
\right.  \label{eq:Kthetar}
\end{equation}

Equations  (\ref{eq:Ktr}) and  (\ref{eq:Kthetar}) show that the
time-like helixes congruence $\Gamma$ defines a Killing field in
the submanifold $\sigma _{r}\;\subset \mathcal{M}^{4}$ but this is
not
a  Killing field in $\mathcal{M}^{4}$.\\

We get the same result if we express the Killing tensor by using
the Born tensor $\widetilde{K}_{\mu \nu }$ and the curvature
vectors $\widetilde{C}_{\nu }$ of the lines of the congruence
$\Gamma \,$:
\[
K_{\mu \nu }\equiv \gamma _{\mu ;\nu }+\gamma _{\nu ;\mu }=\widetilde{K}%
_{\mu \nu }+\gamma _{\mu }\widetilde{C}_{\nu }+\widetilde{C}_{\mu }\gamma
_{\nu }
\]
For the rotating disk ($\widetilde{K}_{\mu \nu }=0$) we simply
obtain:
\begin{equation}
K_{\mu \nu }=\gamma _{\mu }\widetilde{C}_{\nu }+\widetilde{C}_{\mu }\gamma
_{\nu }  \label{bb}
\end{equation}

Equation(\ref{bb}) is very interesting, because it shows the
geometrical meaning of the fact that the Killing tensor is zero in
the submanifold $\sigma _{r}\;\subset \mathcal{M}^{4}$, but it is
not zero in $\mathcal{M}^{4}$. In fact, the congruence $\Gamma$ of
time-like helixes is geodesic on $\sigma _{r}\,$ (where
$\widetilde{C}_{\mu }=0 $), but of course not in
$\mathcal{M}^{4}$,\footnote{Apart from the degenerate case $r=0$,
which corresponds to a straight line in $\mathcal{M}^{4}$} where
the curvature vector $\tilde{C}_{\mu }=\gamma ^{\alpha }\gamma
_{\mu ;\alpha }$ has the following non-null component:
\begin{equation}
\widetilde{C}_{1}=\gamma ^{0}\frac{\frac{\omega ^{2}r}{c^{2}}}{\sqrt{1-\frac{%
\omega ^{2}r^{2}}{c^{2}}}}=\frac{\omega ^{2}r}{c^{2}-\omega
^{2}r^{2}} \label{eq:vettoredicurvatura}
\end{equation}
As a consequence, the Killing tensor has the following non-null
components:

\begin{equation}
K_{01}=\gamma _{0}\widetilde{C}_{1}=\frac{\omega ^{2}r}{c^{2}}\left( \sqrt{1-%
\frac{\omega ^{2}r^{2}}{c^{2}}}\right) ^{-1}  \label{eq:Ktr1}
\end{equation}
\begin{equation}
K_{2 1}=\gamma _{2 }\widetilde{C}_{1}=-\frac{\omega ^{3}r^{3}}{%
c^{3}}\left( \sqrt{1-\frac{\omega ^{2}r^{2}}{c^{2}}}\right) ^{-3}
\label{eq:Kthetar1}
\end{equation}
Equations (\ref{eq:Ktr1}) and  (\ref{eq:Kthetar1}) are in
agreement, respectively, with equations (\ref{eq:Ktr}) and
(\ref{eq:Kthetar}).

\pagebreak


\begin{thebibliography}{99}

\bibitem{ehrenfest}  P.~Ehrenfest, \textit{Phys.~Zeits.}, \textbf{10}, 918 (1909)

\bibitem{sagnac}  M.G. Sagnac, \textit{C.R. Acad. Sci. Paris},
\textbf{157}, 708,1410  (1913)


\bibitem{planck} M. Planck, \textit{Phys. Z.}, \textbf{11}, 294
(1910)

\bibitem{lorentz} H.A. Lorentz, \textit{Nature}, \textbf{106}, 793
(1921)

\bibitem{eddington} A.S. Eddington, \textit{Mathematical Theory of
Relativity}, Cambridge University Press, Cambridge, Eng., 1922, \
\ \textit{Space, Time and Gravitation}, Cambridge University
Press, Cambridge, Eng., 1920

\bibitem{clark} G.L. Clark, \textit{Prc. R. Soc. Edinburgh},
\textbf{62A}, 434 (1947)

\bibitem{berenda} C.W. Berenda, \textit{Phys. Rev.}, \textbf{62},
280 (1942)

\bibitem{cavalleri}  G. Cavalleri, \textit{Il Nuovo Cimento}, LIII B n.2,
416 (1968)

\bibitem{phipps1} T.E. Phipps Jr., \textit{Experiment on Relativistic Rigidity
of a Rotating Disk} NOLTR 73-9 (Naval Ordnance Laboratory, 1973),
p. 47

\bibitem{brotas} A. Brotas, \textit{C.R. Acad. Sci. Paris},
\textbf{267}, 57 (1968)

\bibitem{mccrea}  W.H. McCrea, \textit{Nature},
\textbf{234}, 399 (1971)

\bibitem{einstein} A. Einstein, \textit{The Meaning of
Relativity}, Princeton University Press, Princeton N.J., 1950

\bibitem{stachel} J. Stachel in \textit{General Relativity and
Gravitation},  ed. A. Held, Plenum Press, New York and London,
1980

\bibitem{stedon} G. Stead and H. Donaldson, \textit{Phil. Mag.}, \textbf{20},
92 (1910)

\bibitem{ives} H.E. Ives, \textit{Journ. Opt. Soc. Am.},
\textbf{29}, 472 (1939)

\bibitem{eagle} A. Eagle, \textit{Phil. Mag.}, \textbf{28}, 592
 (1939)

\bibitem{galli} M. Galli, \textit{Rend. Acc. Lincei}, \textbf{12},
569 (1952)

\bibitem{hill}  E.L. Hill, \textit{Phys. Rev.},
\textbf{69}, 488 (1946)

\bibitem{rosen}  N. Rosen, \textit{Phys. Rev.}, \textbf{71}, 54 (1947)


\bibitem{arzelies} H.Arzeli\`{e}s, \textit{Relativistic
Kinematics}, Pergamon Press, New York, 1966

\bibitem{landau}  L. D. Landau and E. M. Lifshitz, \textit{The Classical Theory
of Fields}, Pergamon Press, New York, 1971

\bibitem{moller}  C. M\"{o}ller, \textit{The theory of Relativity}, Oxford
University Press, Oxford, 1972

\bibitem{gron1}  \O . Gr\o n, \textit{Found. Phys.},
\textbf{9}, 353 (1979)

\bibitem{gron2}  \O . Gr\o n, \textit{Int. J. Theor.
Phys.}, \textbf{16}, 603 (1977)

\bibitem{gron3}  \O . Gr\o n, \textit{Am. J. Phys.},
\textbf{43}, 869 (1975)



\bibitem{weber1}  T.A. Weber, \textit{Am. J. Phys},
\textbf{65}, 946 (1997)

\bibitem{dieks}  D.~Dieks, \textit{Eur.~J.~Phys.}, \textbf{12}, 253 (1991)

\bibitem{anandan}  J.~Anandan, \textit{Phys.~Rev.~D}, \textbf{24}, 338 (1981)

\bibitem{rizzi1}  G.~Rizzi and A.~Tartaglia, \textit{Found.~Phys.},
\textbf{28} 1663 (1998)


\bibitem{bergia} S. Bergia and M. Guidone \textit{Found.~Phys.~Lett.},
\textbf{11}, 549  (1998)



\bibitem{selleri1} F. Selleri, \textit{Found. Phys.},
\textbf{26}, 641 (1996)

\bibitem{selleri2} F. Selleri, \textit{Found. Phys. Lett.},
\textbf{10}, 73  (1997)

\bibitem{croca} J. Croca and F. Selleri, \textit{Nuovo Cimento B},
\textbf{114}, 447  (1999)

\bibitem{goy} F. Goy and F. Selleri , \textit{Found. Phys. Lett.},
\textbf{10}, 17  (1997)

\bibitem{vigier} J.P. Vigier , \textit{Phys. Lett. A},
\textbf{234}, 75  (1997)

\bibitem{anas} P.K. Anastasowksi \textit{et al.}, \textit{Found. Phys. Lett},
\textbf{12}, 579  (1999)

\bibitem{rodrigues} W.A. Rodrigues Jr. and M. Sharif, \textit{Found. Phys.},
\textbf{31}, 1767  (2001)



\bibitem{klauber}  R. D. Klauber, \textit{Found. Phys. Lett.}, \textbf{11}
(5), 405 (1998)

\bibitem{klauber2}  R. D. Klauber, \textit{Am. J. Phys.}, \textbf{67} (2),
158 (1999)


\bibitem{tartaglia}  A. Tartaglia, \textit{Found. Phys. Lett},
\textbf{12}, 17  (1999)

\bibitem{grjan} A. Gr\"{u}nbaum and A.J. Janis,
\textit{Synth\`{e}se}, \textbf{34}, 281 (1977)

\bibitem{strauss} M. Strauss, \textit{Int. J. Theor. Phys.},
\textbf{11}, 107 (1974)

\bibitem{cattaneo}  C.~Cattaneo, \textit{Introduzione alla teoria
einsteiniana della gravitazione}, Veschi, Roma, 1961

\bibitem{catt1}  C. Cattaneo,  \textit{Il Nuovo Cimento,} \textbf{10}%
, 318 (1958)

\bibitem{catt2}  C. Cattaneo,  \textit{Il Nuovo Cimento,} \textbf{11}%
, 733 (1959)

\bibitem{catt3}  C. Cattaneo, \textit{Il Nuovo Cimento,} \textbf{13}%
, 237  (1959)

\bibitem{catt4}  C. Cattaneo, \textit{Rend. Acc. Lincei}, \textbf{27}%
, 54  (1959)

\bibitem{born}  M. Born, \textit{Ann. Phys. Lpz.}, \textbf{30}, 1 (1909)

\bibitem{boyer} R.H. Boyer, \textit{Proc. Roy. Soc.},
\textbf{283}, 343 (1965)

\bibitem{pauli} W. Pauli, \textit{Theory of Relativity}, Pergamon
Press, New York, 1958

\bibitem{tangherlini}  F.R. Tangherlini, \textit{Nuovo
Cimento Supp.},  \textbf{20}, 1 (1961)

\bibitem{post}  E. J. Post, \textit{Rev. Mod. Phys.}, \textbf{39 }(2), 475
(1967)

\bibitem{waest}  H.D. Wahlquist and F.B. Estabrook
\textit{J. Math. Phys.}, \textbf{7}, 894 (1966)

\bibitem{wahl}  H.D. Wahlquist, \textit{J. Math. Phys.},
\textbf{33}, 304 (1992)



\bibitem{rizzi2}  G.~Rizzi and A.~Tartaglia,
\textit{Found.~Phys.~Lett.}, \textbf{12}, 179 (1999)

\bibitem{norton}  J. Norton, \textit{Found. Phys.},
\textbf{19}, 1215 (1989)

\bibitem{reichenbach2}  H.~Reichenbach, \textit{The Philosophy of
Space and Time}, Dover, New York, 1957

\bibitem{reichenbach3}  H.~Reichenbach, in \textit{Albert Einstein:
Philosopher-Scientist}, ed. P.A. Schilpp, Evanston, 1949

\bibitem{einstein3}  A.~Einstein, in \textit{Albert Einstein:
Philosopher-Scientist}, ed. P.A. Schilpp, Evanston, 1949


\bibitem{reichenbach}  H.~Reichenbach, \textit{Axiomatization of the Theory of Relativity},
University of California Press, Berkeley and Los Angeles, 1969

\bibitem{Abramowicz}  M. A. Abramowicz, B. Carter and J. P. Lasota, \textit{%
Gen. Rel. Grav., }\textbf{29}, 1173-1183 (1988)

\bibitem{cantoni} V. Cantoni, \textit{Nuovo Cimento B},
\textbf{57} , 220 (1968)

\bibitem{weber}  T.A. Weber, \textit{Am. J. Phys.}, \textbf{67} (2), 159
(1999)

\bibitem{mashh}  B. Mashhoon, \textit{Phys. Lett. A},
\textbf{145},  147 (1990)




\end{thebibliography}
\end{document}